\documentclass[aps,prl,twocolumn,showpacs,amsfonts]{revtex4-1}
\usepackage{color}
\usepackage{graphicx}
\usepackage{epstopdf}
\usepackage{hyperref}
\usepackage{txfonts}

\providecommand{\U}[1]{\protect\rule{.1in}{.1in}}
\providecommand{\U}[1]{\protect\rule{.1in}{.1in}}

\begin{document}

\title{Measuring bipartite quantum correlations of an unknown state}

\author{I. A. Silva$^1$, D. Girolami$^2$, R. Auccaise$^3$, R. S. Sarthour$^4$, I. S. Oliveira$^4$, T. J. Bonagamba$^1$, E. R. deAzevedo$^1$,  D. O. Soares-Pinto$^1$, and G. Adesso$^2$}
\affiliation{$^1\hbox{Instituto de F\'isica de S\~ao Carlos, Universidade de S\~ao Paulo,
Caixa Postal 369, 13560-970 S\~ao Carlos, S\~ao Paulo, Brazil}$ \\
$^2\hbox{School of Mathematical Sciences, The University of Nottingham, University Park, Nottingham NG7 2RD, United Kingdom}$ \\
$^3\hbox{Empresa Brasileira de Pesquisa Agropecu\'aria, Rua Jardim Bot\^anico 1024, 22460-000 Rio de Janeiro, Rio de Janeiro, Brazil}$ \\
$^4\hbox{Centro Brasileiro de Pesquisas F\'isicas, Rua Dr. Xavier Sigaud 150, 22290-180 Rio de Janeiro, Rio de Janeiro, Brazil}$ }

\date{February 28, 2013}
\pacs{03.67.Mn, 03.65.Ta, 03.65.Ud, 03.65.Wj}

\begin{abstract}
We report the experimental measurement of bipartite quantum correlations of an unknown two-qubit state. Using a liquid state Nuclear Magnetic Resonance (NMR) setup and employing geometric discord, we evaluate the quantum correlations of a state without resorting to prior knowledge of its density matrix. The method is applicable to any $2\otimes d$ system and provides,  in terms of number of measurements required, an advantage over full state tomography scaling with the dimension $d$ of the unmeasured subsystem. The negativity of quantumness is measured as well for reference. We also observe the phenomenon of  sudden transition of quantum correlations when local phase and amplitude damping channels are applied to the state.
\end{abstract}

\maketitle

{\it Introduction.}---Quantum Mechanics rules the physical world, but sometimes its effects can be elusive. In particular,  entanglement \cite{entSC}, the genuinely quantum type of correlation shared by two parts of a composite system, is not directly measurable in laboratory. Indeed, there is not a self-adjoint operator quantifying the amount of entanglement of  a state. Hence, an {\it a priori} knowledge of the density matrix appears necessary to evaluate entanglement. This is a serious drawback, since dealing with high dimensional systems makes the state reconstruction extremely demanding if not unfeasible in terms of required resources. The problem has been overcome by introducing nontrivial lower bounds to entanglement measures, expressed as nonlinear functions of the density matrix coefficients, whose values can be detected by means of a limited number of measurements \cite{mint1,mintexp}.

However, it has been acknowledged that even separable (unentangled) mixed states show a distinctive nonclassical behavior, benchmarked by their {\it quantum discord} \cite{OZ,HV}. Arguably considered the most general form of quantum correlations,  it quantifies the minimum induced disturbance on a state of a bipartite system when a local measurement is made on one subsystem. Currently, massive theoretical and experimental efforts are dedicated to discover the potential of such correlations for quantum information processing \cite{rev,datta,white,actistre,activation,rspnature,guexp,cinesinatcomm,gauss,merali}.  The reason is that discord has appealing properties: For pure states, it reduces to entanglement; while for mixed states, even unentangled states (apart from a subset of null measure) have nonvanishing discord  \cite{acinferraro}. A number of phenomenological investigations have proven discord to be easier to create and far more robust than entanglement under decoherent dynamics \cite{disc}.  In spite of that,  even if at theoretical level  a plethora of discord-like measures have been introduced  \cite{rev,rev2},  and some discord witnesses have been implemented \cite{cinesiwitness,mazi1,mazi2}, the situation is similar to entanglement: There is not an observable quantity  grasping the amount of such general quantum correlations in a state.  One could try to recast a measure as a function of density matrix elements and therefore of observable quantities, but most known quantifiers are entropic, thus cannot be easily associated to observables. Taking into account distance-based measures offers a solution to such conundrum.

In this Letter, following the theoretical proposal of Ref.~\cite{pirla}, we report the experimental measurement of quantum correlations of unknown two-qubit states in a room-temperature NMR system \cite{abra,ernst,ivan,brazirev}. In this setting, the information is stored in the nuclear spins, while transformations and state preparation are implemented by applying highly controllable radiofrequency (rf) pulses, magnetic field gradients and evolutions under spin interactions \cite{vandersypen}. On the other hand, the environment affects the spin system by inducing relaxation that drives the system back to the thermal equilibrium distribution according to two independent characteristic times $T_1$, $T_2$ \cite{ivan}.

We pick primarily the geometric discord $D_G$ as quantifier of bipartite quantum correlations \cite{dakic}. Its peculiarity is that it can be expressed, for arbitrary states of $2\otimes d$ systems, as a state-independent function of the density matrix elements, and consequently in terms of observables \cite{pirla,cina}. In fact, there is no need to have full information on the state in order to measure geometric discord. We experimentally verify the theoretical results by preparing an unknown two-qubit Bell diagonal state (the choice of the state is due to experimental convenience exclusively) and retrieving the value of geometric discord by means of local measurements over one of the subsystems. Then, we investigate the robustness of quantum correlations under phase damping and amplitude damping channels acting separately on each qubit. It is predicted that, by appropriately engineering the initial state, measures of discord should undergo a {\it sudden transition} in their dynamical evolution  \cite{mazzola,scnmr}, exhibiting different regimes of resilience to decoherence under noisy conditions. We carry out a comprehensive analysis of general quantum correlations. We monitor in fact the evolution of geometric discord \cite{dakic}, of its lower bound defined in \cite{pirla},  as well as of the so-called negativity of quantumness \cite{taka} for the produced states (the latter measure, here investigated for the first time in open systems), linking the singularity in their expression with the discontinuity in the dynamics.

It is important to discuss the consistency of geometric discord $D_G$ as a reliable estimator of quantum correlations. Recent works have shown that measures built on the Hilbert-Schmidt norm, as geometric discord is, can increase under local reversible operations on the unmeasured subsystem, being biased by the purity of the global state \cite{marco,hybrid}.  In spite of that, geometric discord is still a useful signature of quantum correlations in a number of relevant cases.  
 For example, it can be legitimately used to investigate correlations between system and environment in open quantum evolutions \cite{breuernuovo}, which are globally unitary thus leaving the purity unchanged. Even for a bipartite state under local decoherent evolutions, as studied in this Letter, the geometric discord reliably identifies the sudden transition point in the dynamics as well as other full fledged measures of discord \cite{OZ,mazzola,scnmr}. Specifically, for a system of two qubits (keeping the dimension fixed), as in our case, the problem highlighted in \cite{marco} cannot occur, and $D_G$ admits moreover an operational interpretation in terms of fidelity of remote state preparation \cite{rspnature}, while also bounding from above \cite{interplay} the entanglement quantified by the negativity \cite{negativity}. In general, $D_G$ is a valid lower bound to measures of quantum correlations based on relative entropy \cite{rev}.

 As anticipated, we also investigate experimentally two other discord-like quantities. First, we take in account a nontrivial experimentally-friendly lower bound $Q$ to geometric discord \cite{pirla}: such a quantity is less accurate and does not reveal any of the mentioned dynamical phenomena, as its expression is continuous \cite{nostrareview}. Then, we consider as a reference a fully {\it bona fide} measure of discord, the negativity of quantumness $Q_N^A$ introduced in \cite{activation} and discussed in detail in \cite{taka}. Such a measure quantifies the minimum negativity \cite{negativity} created with an apparatus during a local measurement over one subsystem of a bipartite system. It can be alternatively interpreted (when the measured subsystem is a qubit) in terms of minimum trace distance from the set of zero-discord states.  The $Q_N^A$ is unfortunately harder to compute and less accessible experimentally; to measure it from direct data, without tomography, we need a partial knowledge of the form of the state, specifically the fact that it is a Bell diagonal state. Such a measure is found to detect the same dynamical features as geometric and entropic discord (see also \cite{Saro}).

{\it Theoretical predictions.}---In the following, we consider a bipartite state of a $2\otimes d$ system,  described by a density matrix written in the Bloch representation \cite{bloch}:
\begin{eqnarray*}
\rho &=& \frac14 \sum_{\nu,\lambda=0}^{3, d^2-1} R_{\nu\lambda}\sigma_{\nu} \otimes \tau_{\lambda} \\
&=& \frac 14 \left(\mathbb{I}_{2d}+\sum_{\nu=1}^3 x_{\nu} \sigma_{\nu} \otimes  \mathbb{I}_{d} +\sum_{\lambda=1}^{d^2-1} y_{\lambda} \mathbb{I}_{2}\otimes \tau_{\lambda}  \sum_{\nu,\lambda=1}^{3,d^2-1} c_{\nu\lambda} \sigma_{\nu}\otimes\tau_{\lambda}\right)
\end{eqnarray*}
where $R_{\nu\lambda}=\text{Tr}[\rho(\sigma_{\nu}\otimes \sigma_{\lambda})], \sigma_0=\mathbb{I}_{2}, \sigma_{\nu}$ ($\nu=1,2,3$) are the Pauli matrices and $\{\tau_{\lambda}\}$ their $d$-dimensional generalizations, while $\vec{x}=\{x_{\nu}\},\vec{y}=\{y_{\lambda}\}$ are the column vectors associated to each subsystem, and  $C = (c_{\nu\lambda})$  is the correlation matrix. Quantum discord evaluates the minimum  induced disturbance on the state by a measurement over one subsystem, say $A$. One can adopt a geometric point of view and quantify quantum correlations as the minimum distance of the state to the set of classical-quantum states, i.e., the zero-discord states, which take the form $\chi=\sum_i p_i |i\rangle\langle i|_A\otimes \rho_{Bi}$, where $\sum_i p_i=1$ and $\{|i\rangle\}$ is an  orthonormal vector set \cite{rev}. In this context, the geometric discord introduced in Ref.~\cite{dakic} is defined as $D_G(\rho)=2 \min_{\chi}  \|\rho -\chi \|_2^2$, where the squared Hilbert-Schmidt norm $\|M\|^2_2=\text{Tr}[M^\dagger M]$ is picked as metric. An explicit formula of geometric discord for $2\otimes d$ states is $D_G(\rho)= \frac 12(|\vec x|^2 + \|C\|_2^2 -4 k_{\text{max}})=2 (\text{Tr}[S]-k_{\text{max}})$, where $k_{\max}$ is the largest eigenvalue of the matrix $S = \frac 1{2d}(\vec x {\vec  x}^{\sf T}+  C C^{\sf T})$. Expressing $k_{\max}$  explicitly, the geometric discord can be recast as a state-independent function of the density matrix elements \cite{pirla}:
\begin{eqnarray}\label{dg}
D_G &=&\frac43\text{Tr}[S]-\frac23\sqrt{6\text{Tr}[S^2]-2\text{Tr}[S]^2}\cos\left(\frac\theta3\right),\
\end{eqnarray}
where $\theta =\arccos\big\{\sqrt2\big(2 \text{Tr}[S]^3-9\text{Tr}[S]\text{Tr}[S^2]+9\text{Tr}[S^3]\big)\big(3\text{Tr}[S^2]-\text{Tr}[S]^2\big)^{-\frac32}\big\}$. Fixing $\theta=0$, one obtains a tight lower bound $Q$ to the geometric discord, which is itself a faithful estimator of quantum correlations \cite{pirla}. The result allows to recast $D_G$ and $Q$ as functions of the expectation values of a set of observables $\{\mathcal{O}_i\}$: $D_G=f_{D_G}[\langle \mathcal{O}_i\rangle], Q=f_{Q}[\langle \mathcal{O}_i\rangle]$ \cite{nostrareview}.
The choice of the specific operators depends on the experimental setting. A proposal for an all-optical setup  has been advanced in Refs. \cite{pirla,cina}. In the present Letter, an NMR system is used. In this context, the column vectors $\vec{x}$ and $\vec{y}$ are proportional to the magnetization of each nuclear spin, $\vec{x} = 2\langle \vec{I}\otimes \mathbb{I}_{2}\rangle$ and $\vec{y} = 2\langle\mathbb{I}_{2}\otimes\vec{I}\,\rangle$, where $\vec{I} = \{I_{x}, I_{y}, I_{z}\}$ is the nuclear spin operator, which for spins -- $\frac12$ is $\vec{I} = \vec{\sigma}/2$. The elements of the correlation matrix $c_{\nu \lambda} = \langle \sigma_{\nu}\otimes \sigma_{\lambda} \rangle = 4\langle I_{\nu}\otimes I_{\lambda} \rangle$ and  $\vec{x} = 2\langle \vec{I}\otimes \mathbb{I}_{2}\rangle$ are proportional to a combination of multi quantum coherences \cite{ernst,ivan}.

A full state reconstruction  demands $4d^2-1$ spin measurements. However, to compute $D_G$ and $Q$ one can get rid of $d^2-1$ local measurements on one the subsystems, as the Bloch vector $\vec{y}$ does not account for the $S$ matrix. Indeed, quantum correlations are evaluated from observables of the type $\langle \mathcal{O}_{\nu\lambda}\rangle = \text{Tr}[\sigma_{\nu}\otimes\tau_{\lambda}\rho], \nu=1,\ldots,3; \lambda=0,\ldots, d^2-1$, but global measurements can be replaced by local ones \cite{brazirev}:
\begin{eqnarray}\label{rotation}
\text{Tr}[(\sigma_{\nu}\otimes\sigma_{\lambda})\rho] &=& \text{Tr}[(\sigma_{1}\otimes \mathbb{I}_{d})\xi_{\nu\lambda}] \nonumber \\
\xi_{\nu\lambda} &=& U_{\nu\lambda}\rho U_{\nu\lambda}^{\dagger},
\end{eqnarray}
where $U_{\nu\lambda} = K_{A\rightarrow B}R_{\phi_{\nu}, \phi_{\lambda}}(\theta_{\nu\lambda})$, for $R_{\phi_{\nu}, \phi_{\lambda}}(\theta_{\nu\lambda}) = R_{\phi_{\nu}}^{A}(\theta_{\nu\lambda})\otimes R_{\phi_{\lambda}}^{B}(\theta_{\nu\lambda})$, being $R_{\phi_{\nu(\lambda)}}^{A(B)}(\theta_{\nu\lambda})$ a rotation by an angle $\theta_{\nu\lambda}$ over the direction $\phi_{\nu(\lambda)}$, the indexes $\nu, \lambda = 1, 2, 3$ refer to the rotations for measuring the $C$ matrix elements and $K_{A\rightarrow B}$ represents the CNOT gate with $A$ being the control qubit (see the Supplemental Material \cite{supp} for details).

The negativity of quantumness $Q_N^A$ is defined for $2 \otimes d$ systems as the minimum trace distance from the set of classical-quantum states (see \cite{taka} for other definitions and interpretations), $Q_N^A(\rho)=\frac12 \min_{\chi}  \|\rho -\chi \|_1$, where $\|M\|_1=\text{Tr}\big[\sqrt{M^\dagger M}\big]$ is the trace norm. For the particular case of Bell diagonal two-qubit states, $Q_N^A$ is analytically computable \cite{taka,giovannetti}. Denoting the ordered singular values of the Bloch correlation matrix $C$ as $|c_i| \geq |c_j| \geq |c_k|$, where $i,j,k$ are permutations of $1,2,3$, then the negativity of quantumness is given by half the intermediate one, $Q_N^A=|c_j|/2$. Therefore the negativity of quantumness can be also experimentally determined  as described above.

{\it Experimental results.}---In our implementation we set $d=2$. The state of a NMR two-qubit system in the high temperature approximation is given by $\rho=\frac{1}{4}\mathbb{I}_{4}+\varepsilon\Delta\rho$, where $\varepsilon=\hbar\omega_L/4k_BT\sim10^{-5}$ is the ratio between the magnetic and thermal energies, $\omega_L$ is the Larmor frequency, $k_B$ is the Boltzmann constant and $T$ the room temperature \cite{abra}. All measurements and transformations affect only the deviation matrix $\Delta\rho$, which in fact contains the information about the system state. The unitary operations over $\Delta\rho$ are implemented by radiofrequency pulses and evolutions under spin interactions with an excellent control of the rotation angle and direction. NMR probes the  transverse magnetization, which is proportional to the average values of the  $\langle I_{x} \rangle$ or $\langle I_{y} \rangle$ operators, so only a few elements (single quantum coherences) are directly accessible. In this direction, to obtain a full characterization of $\Delta\rho$ it is necessary to execute a set of independent measurements after applying specific rotations to the system, which characterize NMR quantum state tomography \cite{ivan}. It is worth remarking that, since in NMR experiments only the deviation matrix is detected, the calculations of the $\vec{x}$ vector and of the correlation matrix $C$ are done in units of $\varepsilon$.

\begin{figure}[h!]
\begin{center}
\includegraphics[scale=0.6]{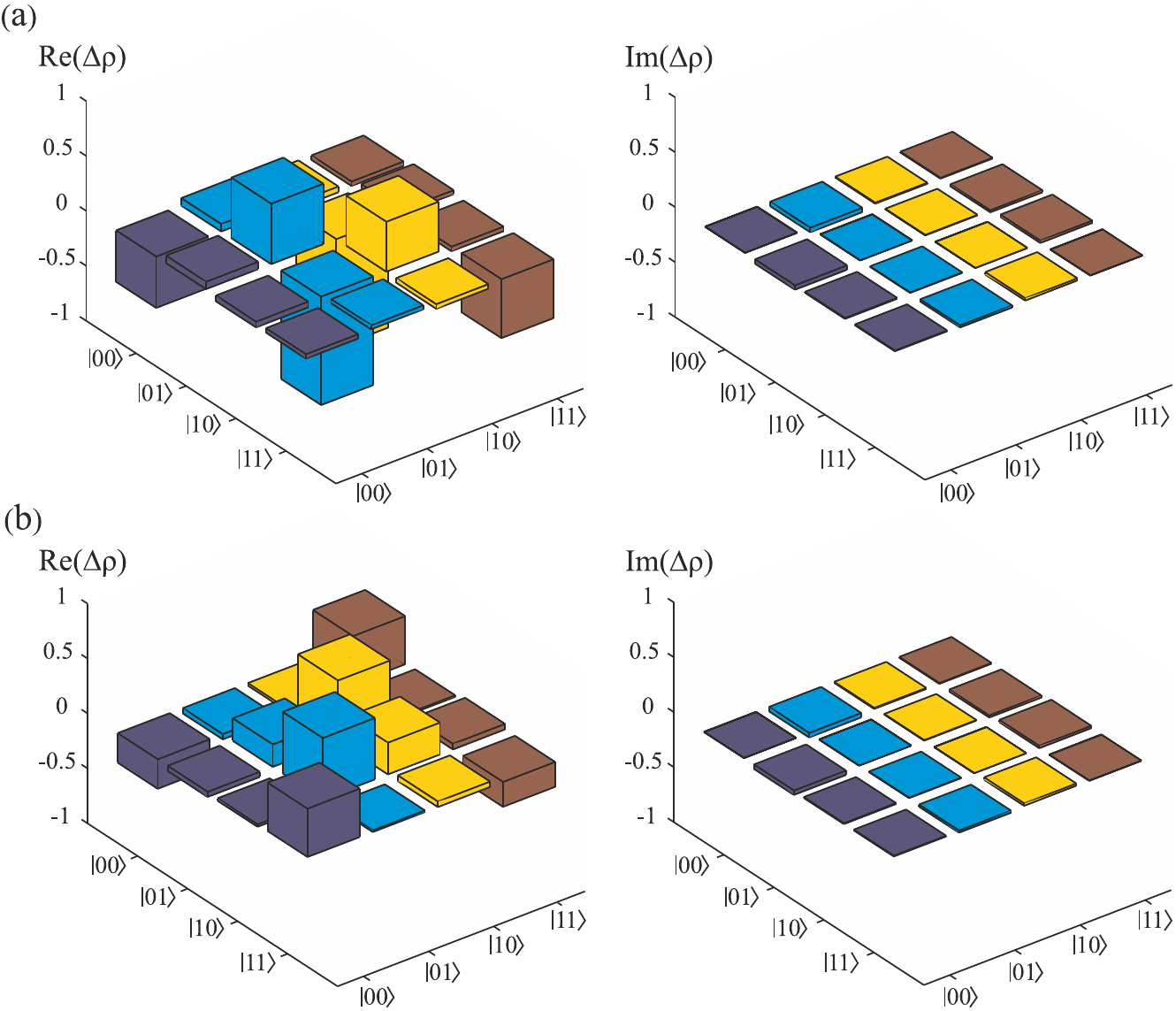}
\caption{(color online).\label{initial_state} Block diagrams for the deviation density matrix related to the Bell diagonal initial states. For $\rho_1$ in Fig.\ref{initial_state}(a) we have $|c^{(1)}_1| = |c^{(1)}_2| = |c^{(1)}_3| = 0.2$, while $\rho_2$  in Fig.\ref{initial_state}(b) reads $|c^{(2)}_1| = 0.5$, $|c^{(2)}_2| = 0.06$, and $|c^{(2)}_3| = 0.24$.}
\end{center}
\end{figure}

The experiments were performed on a liquid state carbon-13 enriched chloroform sample (CHCl$_{3}$) at room temperature, with the two qubits being encoded in the $^{1}$H and $^{13}$C spin--$\frac12$ nuclei \cite{supp}. Two initial Bell diagonal states $\rho_{1,2}=\frac 14(\mathbb{I}_4+\sum_i c^{(1),(2)}_i \sigma_i\otimes \sigma_i)$  were prepared by mapping the correlation matrix into the deviation matrix, as described in  \cite{scnmr,supp}. Note that the methodology here employed is not restricted to this class of states, but extends to arbitrary states of $2\otimes d$ systems. To verify the correct preparation of the states, the pulse sequences proposed in Refs. \cite{mazi1,scnmr} were applied followed by a quantum state tomography procedure \cite{tomo}.  The resulting deviation matrices are shown as block diagrams in Fig. \ref{initial_state}. We remark that tomography is performed just to test the quality of the state preparation procedure, and the acquired information is not used for the direct estimation of quantum correlations. The interaction with the environment in a NMR system is described by phase and amplitude damping channels \cite{nielsen_chuang}. As can be seen in Refs. \cite{mazi1,scnmr}, the correlations presented in the initial state of Fig.\ref{initial_state}(a) should decay monotonically in time, while the state of Fig.\ref{initial_state}(b) is expected to exhibit a nontrivial behavior of quantum correlations during the evolution.

According to Eq.~(\ref{dg}) and the accompanying discussion, quantum correlations in a two-qubit system are obtained from the expectation values of the correlation matrix elements  $c_{\nu \lambda} = \langle \sigma_{\nu}\otimes \sigma_{\lambda} \rangle =\text{Tr}[(\sigma_{\nu} \otimes \sigma_{\lambda})\rho]$ and  $\vec{x} = \langle \vec{\sigma}\otimes \mathbb{I}_{2}\rangle$. These can be calculated for instance from the experimentally tomographed deviation matrix. However, using the rotations described by Eq.~(\ref{rotation}) with proper set of angles, the evaluation of  $D_G$ and $Q$ is reduced to a set of  spin magnetization measurements on one of the qubits. This means that, after a correct set of rotations (rf pulses) applied to the prepared state, the  geometric discord (and its lower bound) can be determined directly from the NMR signals (apart from the $\varepsilon^2$ factor), without having to know the state. We dub this procedure as {\it direct measurement}. The procedure is repeated until the thermal equilibrium state is re-established.   Concerning the negativity of quantumness \cite{taka}, a partial knowledge of the state is instead required to evaluate its value. In particular, assuming (as {\it a posteriori} verified by tomography, see \cite{supp}) that the state remains in Bell diagonal form during the evolution, then the direct method still suffices to extract the correct value of $Q_N^A$.  To benchmark the effectiveness of the direct method, overall, we compared its outcomes with the results obtained for the corresponding measures of quantum correlations by evaluating them on the state reconstructed by complete tomography as well.

\begin{figure}[t]
\begin{center}
\includegraphics[scale=0.5]{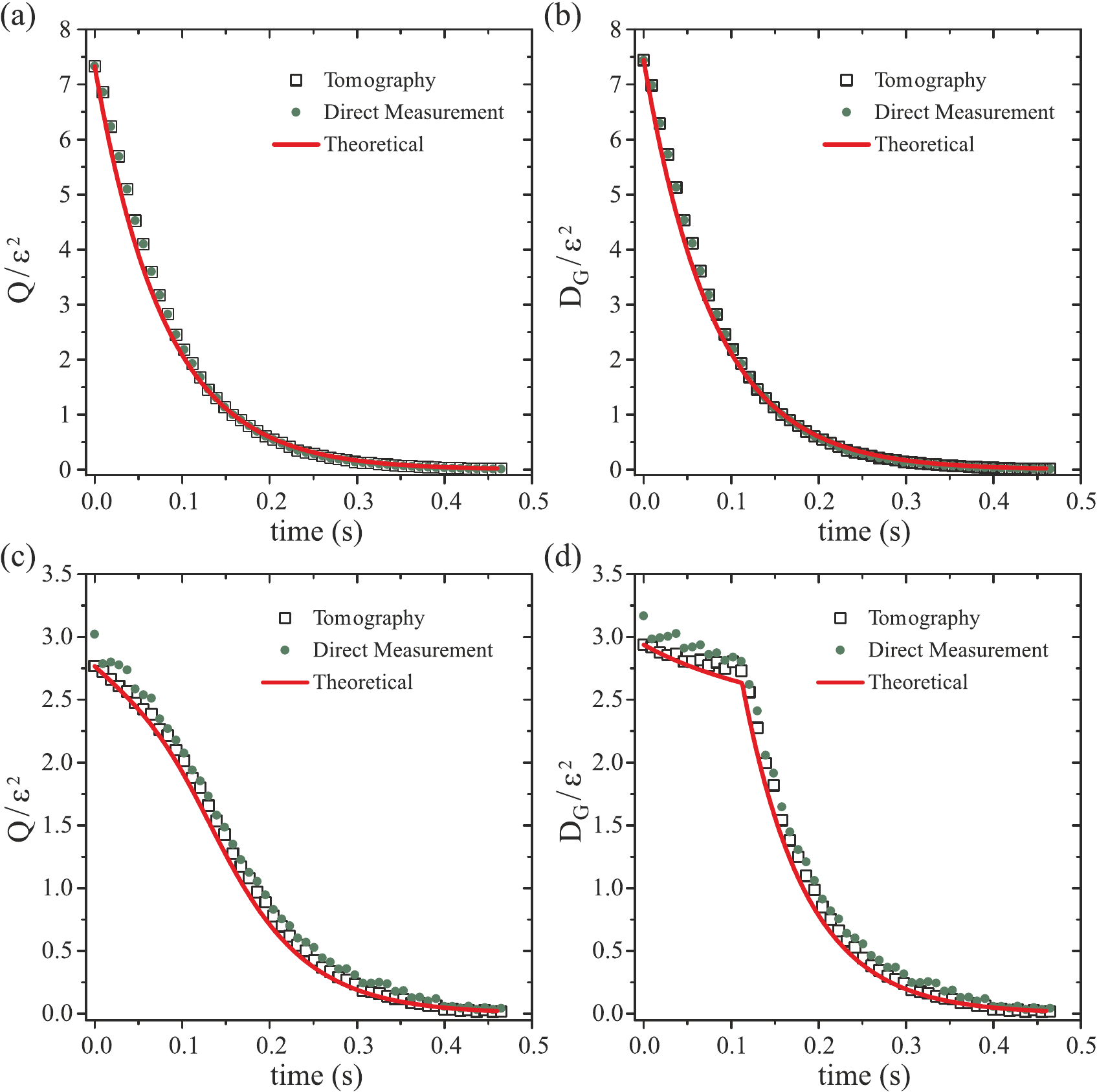}
\caption{(color online). \label{graphics}Time evolution of $Q$ and $D_G$ for the state $\rho_1$ in (a) and (b), and $\rho_2$ in (c) and (d). The green dots represent the direct measurement and the black open squares corresponds to tomography results. The red lines depict the theoretical predictions. The relaxation times retrieved from experimental data  are:   $T_1 = 3.57\,s$ and $T_2 = 1.2\,s$ for the hydrogen, and $T_1 = 10\,s$ and $T_2 = 0.19\,s$ for the carbon. Quantum  correlations are displayed in units of $\varepsilon^2$.}
\end{center}
\end{figure}

\begin{figure}[t]
\begin{center}
\includegraphics[scale=0.5]{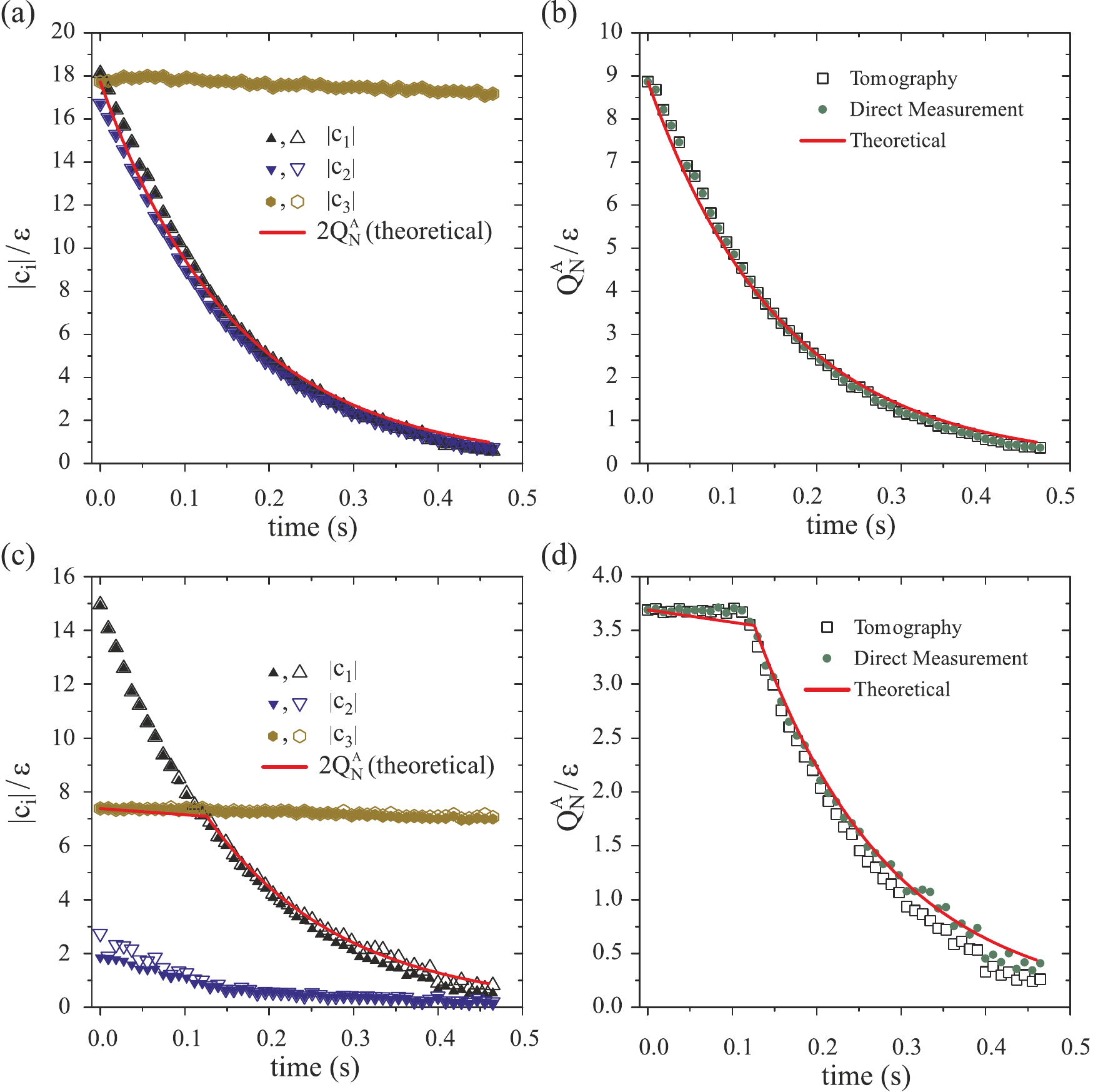}
\caption{(color online).\label{negativity} Time evolution of $|c_1|$ (upward triangle), $|c_2|$ (downward triangle) and $|c_3|$ (hexagon) for state $\rho_1$ in (a), and $\rho_2$ in (c). The open symbols represent the tomography results and the filled symbols represent the direct measurement. Note that for each $|c_i|\,(i = 1, 2, 3)$ both forms of measurement are superimposed. In (b), for the initial state $\rho_1$, and (d), for the initial state $\rho_2$, the green dots represent the direct measurement and the black squares represent the tomography results. The red lines are the theoretical predictions for $Q_{N}^{A}$. Quantum correlations are displayed in units of $\varepsilon$.}
\end{center}
\end{figure}

The results obtained for  $D_G$ and $Q$ using the direct measurement and tomography procedures as well as the  theoretical predictions are reported in Fig.~\ref{graphics} for both experimentally produced states. In Fig.\ref{negativity}, we present the results obtained for $Q_N^A$ by direct measurements  (the $^{1}H$ nucleus was detected), tomography and theoretical predictions for both states.
We highlight a satisfactory agreement between the direct measurements, the tomographic data, and the theoretical predictions. This demonstrates that we are able to directly quantify bipartite quantum correlations in unknown (or partially known in the case of $Q_N^A$) two-qubit states with our NMR setup. In particular, for the first time the negativity of quantumness is observed to undergo sudden transition in the same dynamical conditions as the geometric discord and the entropic discord \cite{mazzola,Saro}. Furthermore, directly assessing the value of $D_G$ provides a nontomographic method to estimate {\it a priori} the suitability of the produced states for quantum communication via remote state preparation \cite{rspnature}.

{\it Conclusions.}--- NMR systems are a natural arena for quantum information processing with negligible entanglement, thus ideal testbeds  for investigating dynamical properties of quantum correlations in  open system dynamics. In this respect, the information stored in the density matrix of a state is often redundant to the quantification of its quantum correlations. Recent theoretical findings allow one to implement optimized protocols  by reducing the number of required measurements as compared to the full state reconstruction. Here we have considered an NMR system and detected the quantum correlations of a two-qubit Bell diagonal state.
Furthermore, we observed the sudden transition of geometric discord and negativity of quantumness under phase and amplitude damping channels. The sudden transition and in particular the {\it freezing} \cite{mazzola}, common to various measures of quantum correlations other than entanglement under particular decoherent evolutions, is certainly worthy of further theoretical and experimental investigation \cite{ben}.

We finally stress that the reported experiment has to be taken as a {\it proof of principle}. Thanks to the explicit closed formula, without any knowledge of the initial state of a $2\otimes d$ system, it is possible to evaluate geometric discord by $3d^2$ local spin measurements, and tune the dynamics in order to protect quantum correlations from decoherence. Also, it is remarkable that an optical setup would allow to detect quantumness of correlations by only seven projective measurements for any $d$ \cite{pirla}, but the implementation of such quantum operations appears challenging even with state-of-the-art technology \cite{nostrareview}.

\emph{Acknowledgments.}---We thank C.-F. Li, R. Lo Franco, O. Moussa, G. Passante, and M. Piani for fruitful discussions. We acknowledge financial support from the UK EPSRC Nottingham Research Development Fund [Grant PP-0313/36] and the Brazilian funding agencies CNPq, CAPES [Pesquisador Visitante Especial - Grant No. 108/2012], FAPERJ, and FAPESP [Grants 2009/18354-8 (ERdeA), 2009/54880-6 and 2012/02208-5 (TJB)]. This work was performed as part of the Brazilian National Institute of Science and Technology for Quantum Information (INCT-IQ).

\clearpage

\appendix*
\section{SUPPLEMENTAL MATERIAL}
\setcounter{equation}{0}

\section{\textbf{NMR experiments}}

The NMR experiments were performed on a carbon-13 enriched chloroform sample (CHCl$_{3}$), where the two nuclear spins of $^{1}$H and $^{13}$C implement the qubits. The sample was prepared by mixing $100$ mg of $99$ \% $^{13}$C-labelled CHCl$_{3}$ in $0.7$ mL of $99.8\%$ CDCl$_{3}$ in a $5$ mm NMR tube (both compounds provided by the Cambridge Isotope Laboratories Inc.). NMR experiments were carried out at $25^{\circ}$C in a Varian 500 MHz Premium Shielded spectrometer located at the Brazilian Center for Research in Physics (CBPF, Rio de Janeiro). A Varian $5$ mm double resonance probehead equipped with a magnetic field gradient coil was used. Two $\pi/2$ pulses of length $7.4$ $\mu$s and $8.0$ $\mu$s were applied to $^{1}$H and $^{13}$C, respectively. Spin-lattice relaxation times (T$_{1}$) for $^{1}$H and $^{13}$C nuclei, measured by the inversion recovery pulse sequence, were $3.57$ s and $10$ s. Spin-spin relaxation times (T$_{2}$), measured by a CPMG pulse sequence, were estimated to be about $1.2$ s and $0.19$ s for $^{1}$H and $^{13}$C nuclei. The recycle delay (d$_{1}$) was set at $60$ s in all experiments.

The nuclear spin Hamiltonian is given by
\begin{eqnarray}
\mathcal{H} &=&-\left(\omega_{H}-\omega _{rf}^{H}\right) I_{z}^{H}-\left(\omega _{C}-\omega _{rf}^{C}\right) I_{z}^{C}+2\pi J\,I_{z}^{H}\,I_{z}^{C} \\
&+&\omega _{1}^{H}\left(I_{x}^{H}\cos\varphi^{H}+I_{y}^{H}\sin\varphi^{H}\right) +\omega_{1}^{C}\left(I_{x}^{C}\cos\varphi^{C}+I_{y}^{C}\sin\varphi^{C}\right), \nonumber
\label{Hamiltoniano}
\end{eqnarray}
where $I_{\alpha}^{H}\left(I_{\beta}^{C}\right)$ is the spin angular momentum operator in the $\alpha,\beta=x,y,z$ direction for $^{1}$H ($^{13}$C); $\varphi^{H}\left(\varphi^{C}\right)$ defines the direction of the rf field (pulse phase) and $\omega_{1}^{H}\left(\omega_{1}^{C}\right)$ is the rf nutation frequency (RF power) for $^{1}$H ($^{13}$C) nuclei. The first two terms describe the Zeeman interaction between the $^{1}$H and $^{13}$C nuclear spins and the main
magnetic field B$_{0}$. The corresponding frequencies are $\omega_{H}/2\pi\approx 500$ MHz and $\omega_{C}/2\pi\approx125$ MHz. The third term is due to a scalar spin-spin coupling of $J\approx215.1$ Hz. The fourth and fifth terms represent the rf field to be applied to the $^{1}$H and $^{13}$C nuclear spins, respectively. There is also a time-dependent coupling of the nuclear spins with the environment that includes all fluctuating NMR interactions accounting for the spin relaxation, e.g., $^{1}$H - $^{13}$C dipolar spin-spin couplings and interactions with the chlorine nuclei.

To investigate the dynamics of quantum correlations, we prepared a Bell-diagonal state:
\begin{equation}
\rho = \frac 14\left( \begin{array}{cccc}
1+ c_{3} & 0 & 0 & c_{1}-c_{2} \\
0 &1 -c_{3} & c_{1}+c_{2} & 0 \\
0 & c_{1}+c_{2} &1 -c_{3} & 0 \\
c_{1}-c_{2} & 0 & 0 &1+ c_{3} \end{array}\right),
\end{equation}
where the coefficients $\{c_i\}$ are picked as explicited in Fig.2(a) of the main text. Such state is obtained from thermal equilibrium by applying the pulse sequence for producing the pseudo-pure state $\left|11\right\rangle$, followed by rf pulses which implement a pseudo EPR gate \cite{chuang1998PRSLA}. It is worth mentioning that a true EPR gate would require additional pulses. To prepare the state showed in Fig.2(b), a diagonal state was created from the thermal equilibrium by  $5.5$ ms strongly modulated pulses (SMP) -- designed using a MATLAB self-written routine based on a SIMPLEX optimization protocol -- followed by a $2$ ms magnetic field gradient pulse. The deviation matrix is $\Delta\rho=\text{diag}(\alpha, \beta,\gamma,\delta)$, whose entries are defined through the optimized SMP pulse.
Next, the deviation matrix is transformed by a pseudo-EPR gate into an X-type matrix \cite{chuang1998PRSLA}:
\begin{eqnarray}
\Delta\rho &  =\frac{1}{2}\left(
\begin{array}
[c]{cccc}%
\alpha+\gamma & 0 & 0 & -\alpha+\gamma\\
0 & \beta+\delta & -\beta+\delta & 0\\
0 & -\beta+\delta & \beta+\delta & 0\\
-\alpha+\gamma & 0 & 0 & \alpha+\gamma
\end{array}
\right)  \text{.} \label{estadoX}%
\end{eqnarray}
The SMP was designed to produce a diagonal state with populations $\alpha$, $\beta$, $\gamma$ and $\delta$, so that, after the transformation by the pseudo-EPR gate, a state with a deviation matrix in the form equivalent to the state in Eq. (2) of the main text, with specified $c_{i}$'s, was obtained. After having prepared the desired state, the system was left to evolve for a period $\tau=m/4J$, with $m=1,2,3,4,..$. The dynamics, considering on-resonance evolution, was characterized by two terms of the Hamiltonian shown in Eq. (\ref{Hamiltoniano}) acting on the nuclear spins: the scalar coupling and the time-dependent term, which account for the system relaxation.

\begin{figure}[t]
\begin{center}
\includegraphics[scale=0.4]{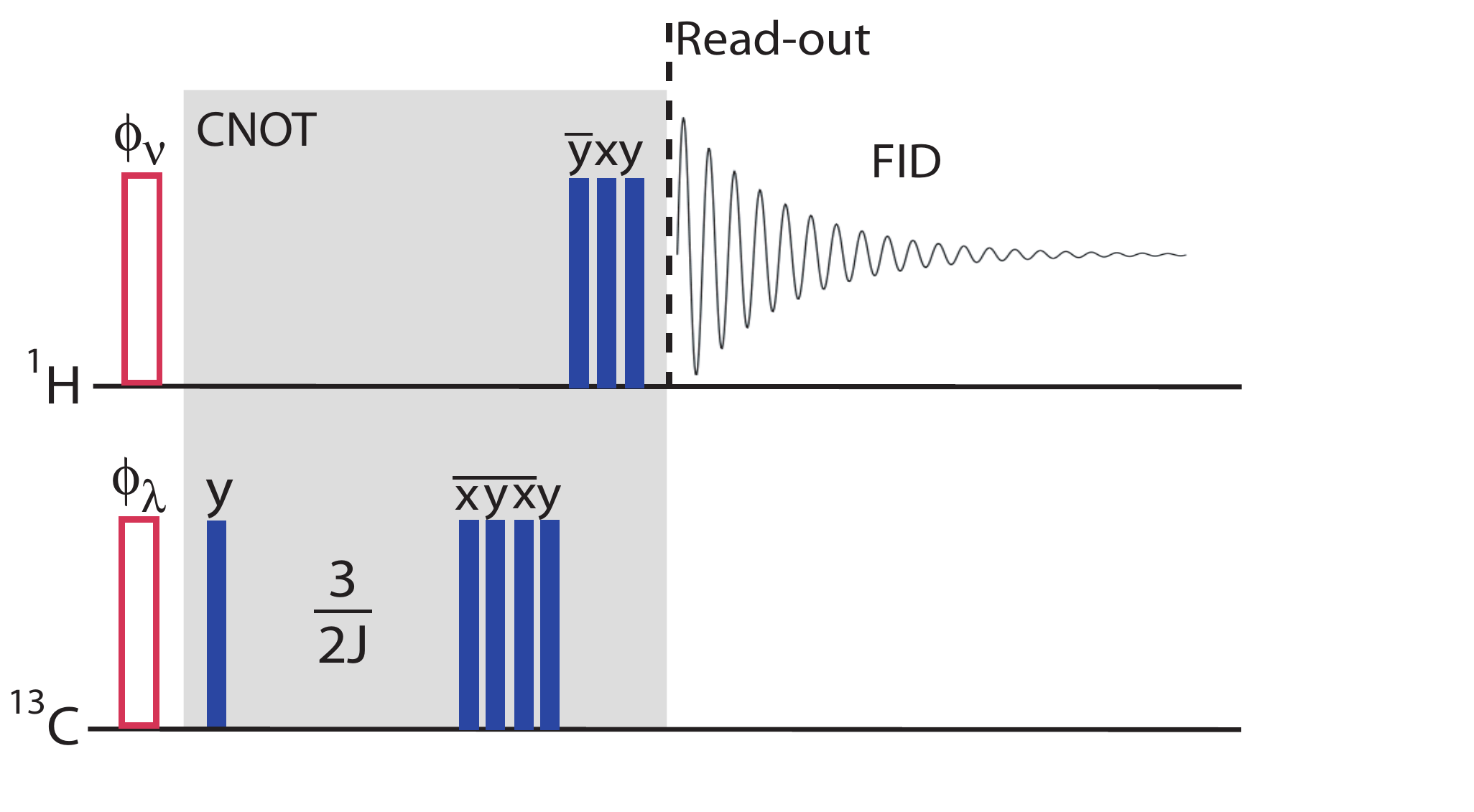}
\caption{\label{pulse_seq}Schematic representation of the pulse sequence for the direct magnetization measurements. The unfilled red bars represent pulses whose phases and angles depend on the indexes $\nu$, $\lambda = 1$, $2$, $3$, which refer to the measure of each element of the $C$ matrix. The phases and angles are $\phi_{1} = x$, $\phi_{2} = z$, $\phi_{3} = y$, $\theta_{11} = 0$, $\theta_{22} = \theta_{33} = \pi/2$, $\theta_{12} = \theta_{21} = 3\pi/2$, $\theta_{13} = \theta_{31} = \pi/2 = \theta_{23} = \theta_{32} = \pi/2$. The thinner blue bars represent a $\pi /2$ pulse realizing the CNOT logical gate.}
\end{center}
\end{figure}

After the evolution period, to read-out the quantum state, we implemented the protocol of Fig.\ref{pulse_seq}. The $y$ rotation was carried out by a single rf pulse, while for the $z$ rotation the pulse sequence $\left(\frac{\pi}{2}\right)_{-y}$-$\left(\frac{\pi}{2}\right)_{x}$-$\left(\frac{\pi}{2}\right)_{y}$ was applied to both nuclei. The controlled-NOT gate was achieved by the pulse sequence $\left(\frac{\pi}{2}\right)_{y}^{C}$-$U\left(\frac{3}{2J}\right)$-$\left(\frac{\pi}{2}\right)_{-x}^{C}$-$\left(\frac{\pi}{2}\right)_{-y}^{C}$-$\left(\frac{\pi}{2}\right)_{-x}^{C}$-$\left(\frac{\pi}{2}\right)_{y}^{C}$-$\left(\frac{\pi}{2}\right)_{-y}^{H}$-$\left(\frac{\pi}{2}\right)_{x}^{H}$-$\left(\frac{\pi}{2}\right)_{y}^{H}$, where the super indices states for nucleus where the pulse is applied and $U\left(\frac{3}{2J}\right)$ represent a free evolution under $J$ coupling. In this case, a pulse sequence implementing a true CNOT gate is indeed necessary to produce the correct output. As described above, the experimental procedure depicted in Fig. 1 was run three times for each initial state in order to measure the magnetization  $\langle I_{1}^{H}\rangle_{\xi_{i}}$ in the states $\xi_{i}$ that leads to determine the two-point correlation functions $\langle I_{i}^{H}\otimes I_{i}^{C}\rangle _{\rho}$. So, the quantum correlations measured by $D_G, Q, Q_N^A$ could be directly evaluated. The rotations to transform all of the elements of the correlation matrix into magnetization ones ($\langle I_{i}^{H}\otimes I_{i}^{C}\rangle _{\rho} \rightarrow \langle I_{1}^{H}\rangle_{\xi_{i}}$) are given by:
\begin{eqnarray*}
R_{\phi_{1}, \phi_{1}} (\theta_{11}) &=& R_{xx}(0) \\
R_{\phi_{2}, \phi_{2}} (\theta_{22}) &=& R_{zz}(\pi/2) \\
R_{\phi_{3}, \phi_{3}} (\theta_{33}) &=& R_{yy}(\pi/2) \\
R_{\phi_{1}, \phi_{2}} (\theta_{12}) &=& R_{xz}(3\pi/2) \\
R_{\phi_{2}, \phi_{1}} (\theta_{21}) &=& R_{zx}(3\pi/2) \\
R_{\phi_{1}, \phi_{3}} (\theta_{13}) &=& R_{xy}(\pi/2) \\
R_{\phi_{3}, \phi_{1}} (\theta_{31}) &=& R_{yx}(\pi/2) \\
R_{\phi_{2}, \phi_{3}} (\theta_{23}) &=& - R_{zy}(\pi/2) \\
R_{\phi_{3}, \phi_{2}} (\theta_{32}) &=& - R_{yz}(\pi/2). \\
\end{eqnarray*}

The quantum state was also read-out  by quantum state tomography, as described in Ref. \cite{long2001,teles}, in order to check the consistency of the results obtained by direct measurements. This process allowed us to reconstruct the deviation density matrix of the system for each instant of time, benchmarking its behavior under the coupling with the environment. Multi-step increments of $\tau$ in successive experiments allowed the effect of the spin environment on the initial deviation matrix to be followed. Final states corresponding to $251$ distinct $\tau$ values were acquired. The high homogeneity of the static magnetic field was guaranteed by the linewidth of $\approx0.95$ Hz in $^{1}$H spectra.

\section{\textbf{Decoherence analysis}}

The decoherence process is theoretically analysed through the operator sum representation
technique \cite{Nielsen}, in which the evolution of the density operator is
given by \cite{qchannels}
\begin{displaymath}
{\cal E}(\rho_{AB}) = \sum_{i,j}(E_{i}^{A}\otimes\mathbb{I}_{B})(\mathbb{I}_{A}\otimes E_{j}^{B})\rho_{AB}(\mathbb{I}_{A}\otimes E_{j}^{B})^{\dagger}(E_{i}^{A}\otimes\mathbb{I}_{B})^{\dagger}  ,
\end{displaymath}
where $E_{k}^{\alpha}\left(t\right)$ are the Kraus operators.

NMR presents, typically, two relaxation channels, namely the generalized amplitude damping and the phase damping.

The amplitude damping channel is described by the following set of operators
\begin{eqnarray}
E_{0} &=&\sqrt{\gamma}%
\left(
\begin{array}{cc}
 1& 0  \\
 0 & \sqrt{1-p}
\end{array}
\right),\quad E_{1}=\sqrt{\gamma}%
\left(
\begin{array}{cc}
 0& \sqrt{p}  \\
 0 & 0
\end{array}
\right),\\
E_{2}  &  =&\sqrt{1-\gamma}%
\left(
\begin{array}{cc}
 \sqrt{1-p}& 0  \\
 0 & 1
\end{array}
\right),\quad E_{3}=\sqrt{1-\gamma}%
\left(
\begin{array}{cc}
 0& 0  \\
 \sqrt{p} & 0
\end{array}
\right),\nonumber
\end{eqnarray}
where, in the NMR context, $\gamma=1/2-\varepsilon/2$ and $p=1-\exp\left(-t/T_{1}\right)$, $T_{1}$ being the longitudinal relaxation time of the qubit under consideration. We observe that in our case, the relaxation times are different for the two qubits, since they have distinct Larmor frequencies.

For the phase damping channel, one has
\begin{eqnarray}
E_{4}=\sqrt{1-\frac{\lambda}{2}}%
\left(
\begin{array}{cc}
 1& 0  \\
 0 & 1
\end{array}
\right)
,\quad E_{5}=\sqrt{\frac{\lambda}{2}}%
\left(
\begin{array}{cc}
 1& 0  \\
 0 & -1
\end{array}
\right),
\end{eqnarray}
where $\lambda=1-\exp\left(-t/T_{2}\right)$ with $T_{2}$ being the transverse relaxation time associated with the qubit.

 The agreement between the model and the experimental data is quite satisfactory.
 In the theoretical model we neglected a number of possible error sources, such as magnetic field inhomogeneities or oscillations of the residual off-diagonal elements of the deviation matrix. In spite of that, all the singular features predicted by the quantum model emerge from the experimental data analysis. Specifically, for Bell-diagonal states one has $S=\frac{C^2}{4}=\text{diag}\{c_1^2/4,c_2^2/4,c_3^2/4\}$ and expressions for $D_G,Q, Q_N^A$ are easily obtained.  While the quantity $Q$ is manifestly continuous, $D_G$ and $Q_N^A$ show discontinuity in the quantum correlations at a critical time.

\end{document}